\documentclass[12pt]{article}
 \pdfoutput=1
\usepackage{jcappub}
\usepackage[top=1in, bottom=1in, left=1in, right=1in]{geometry}
\usepackage[font=footnotesize]{caption}
\usepackage{array}
\usepackage{enumerate}
\usepackage{float}
\usepackage{subfig}
\usepackage{multicol}
\usepackage{multirow}
\usepackage{epsfig}
\usepackage{graphicx}
\usepackage{pict2e}
\usepackage{color}
\usepackage{amsmath,amsfonts,amssymb}
\usepackage{bbm}
\usepackage[utf8]{inputenc}
\usepackage[english]{babel}
\usepackage{xspace}

\newcommand{\be}{\begin{equation}}
\newcommand{\ee}{\end{equation}}
\newcommand{\bea}{\begin{eqnarray}}
\newcommand{\eea}{\end{eqnarray}}

\newcommand{\beq}{\begin{equation}}
\newcommand{\eeq}{\end{equation}}
\newcommand{\cH}{\cal{H}}

\def\d{{\rm d}}
\def\cH{{\cal H}}

\def\vk{\mathbf{k}}

\def\vK{\mathbf{K}}

\title{Constraints on Scalar and Tensor spectra from $N_{eff}$}
\author{Ido Ben-Dayan$^{1}$, Brian Keating$^2$, David Leon$^2$, Ira Wolfson$^{3}$}
\affiliation{$^1$ Physics Department, Ariel University, Ariel 40700, Israel\\
$^2$Department of Physics, University of California San Diego, CA, 92093 USA\\
$^3$Physics Department,
Ben-Gurion University of the Negev, P.O. Box 653, Be'er-Sheva 8410500, Israel}
\emailAdd{ido.bendayan@gmail.com,bkeating@ucsd.edu,d2leon@physics.ucsd.edu,irawolfsonprof@gmail.com}

\begin{abstract}
 {At the linear level, the gravitational wave (GW) spectrum predicted by inflation, and many of its alternatives, can have arbitrarily small amplitude and consequently an unconstrained tilt. However, at second order, tensor fluctuations are sourced by scalar fluctuations that have been measured in the cosmic microwave background (CMB). These second order fluctuations generically produce a minimum amount of tensor perturbations corresponding to a tensor-to-scalar ratio of $r\sim 10^{-6}$. Inverting this relationship yields a bound on the tensor tilt sourced by scalar fluctuations. Since this induced GW spectrum depends on the scalar spectrum,  we derive a new indirect bound that involves \textit{all scales} of the scalar spectrum based on CMB observations. This bound comes from the constraint on the number of effective relativistic degrees of freedom, $N_{eff}$. We estimate the bound using current data, and the improvements expected by future CMB experiment. The bound forces the running and running of running to conform to standard slow-roll predictions of $\alpha,\beta \lesssim (n_s-1)^2$ where $\alpha\equiv \frac{dn_s}{d \ln k}$ and $\beta\equiv \frac{d^2 n_s}{s \ln k^2}$, improving on current CMB measurements by an order of magnitude. This bound has further implications for the possibility of primordial black holes as dark matter candidates. Performing a likelihood analysis, including this new constraint, we find that positive $\alpha$ and/or $\beta$ are disfavored at least at $1\sigma$. Even using conservative analysis $\beta + 0.074\ \alpha>8.6\times 10^{-4}$ are ruled out at $3\sigma$. Finally, using bounds on the fractional energy density of gravitational waves today obtained by LIGO and the Pulsar Timing Array, we obtain a bound on the primordial scalar spectrum on these scales and give forecast for future measurements.}
\end{abstract}

\begin{document}

\maketitle

\section{Introduction}
A stochastic gravitational waves background (GW), could be produced by a multitude of physical phenomena on different eons and scales ranging from Early Universe scenarios through phase transitions to incoherent accumulation of binary black hole coalescence \cite{TheLIGOScientific:2016dpb,Akrami:2018odb,Ade:2018gkx, Aghanim:2018eyx,Bartolo:2016ami,Abazajian:2016yjj}. The fractional energy density stored in GW is therefore an invaluable probe of these physical phenomena. One can probe the energy density of the stochastic gravitational wave background in several ways in different epochs and on different wavelengths.
Cosmic Microwave Background Radiation (CMB) observations, and specifically the B-mode polarization measurements, have persistently probed GW on the largest possible scales and have tested Early Universe scenarios, most notably inflation \cite{Akrami:2018odb,Ade:2018gkx}. Such measurements probe the GW energy density at the time of decoupling. The CMB program is expected to continue in the foreseeable future improving the accuracy of various cosmological parameters by an order of magnitude or more \cite{Ade:2018sbj,Abazajian:2016yjj,Inoue:2016jbg,Chluba:2019kpb}.   More recent measurements on intermediate and small scales include LIGO and the Pulsar Timing Array (PTA) respectively \cite{Kramer:2010tm,TheLIGOScientific:2016dpb}. Such experiments probe the energy density of GW today. While these probes are also sensitive to various Early Universe scenarios, they are mostly expected to detect other physical phenomena such as phase transitions or other astrophysical phenomena, \cite{TheLIGOScientific:2016dpb,Bartolo:2016ami,Kramer:2010tm}.

Early Universe models based on quantum fluctuations, whether inflation or bounce, predict a primordial scalar/density spectrum and a tensor/GW spectrum. Our focus henceforth will be on these spectra and on ways to constrain them. CMB and BAO measurements have measured the scalar spectrum on scales $H_0<k<1 Mpc^{-1}$ to be
\be \label{eq:PSplanck}
P_S=A_s\left(\frac{k}{k_0}\right)^{n_s-1},\quad A_s=2.1\times 10^{-9},\quad n_s=0.965
\ee
and have placed an upper bound on the similarly defined GW spectrum $P_T=A_T\left(\frac{k}{k_0}\right)^{n_T}$ in the form of the scalar to tensor ratio $r$, \cite{Akrami:2018odb,Ade:2018gkx}:
\be \label{eq:rcmb}
r\equiv \frac{P_T}{P_S}|_{k_0}\leq 0.06
\ee
$k_0$ is the so called pivot scale that somewhat differs from one experiment to the other, but for our purposes we shall take it to be $k_0=0.05\; Mpc^{-1}$. In the following, $k_{eq}=0.01\; Mpc^{-1}$ in compliance with Planck 2018 analysis.

Combining LI, PTA and CMB observations, allows us to try and probe not just the amplitude of the GW spectrum $r$, but also its frequency dependence or tilt. Such works have been carried out in e.g. \cite{Boyle:2007zx,Meerburg:2015zua}, and the future LI and CMB measurements will certainly improve these constraints. 

Large parts of the GW spectrum are inaccessible neither to CMB, nor to LI nor PTA experiments in the foreseeable future.
To probe these parts of the spectrum, one resorts to indirect probes.
In the context of GW, \cite{Chluba:2014qia} calculated the predicted spectral distortion signal given a tensor spectrum, while in \cite{Meerburg:2015zua}, a likelihood analysis assuming some GW spectrum $P_T$ was carried out.
Of specific interest is the fact that GW are relativistic degrees of freedom and as such, they affect BBN and the CMB temperature anisotropies measurements. The GW energy density contributes to the number of effective relativistic d.o.f. at the time of decoupling, $N_{eff}$.
The dependence of $N_{eff}$ on the GW spectrum was derived in \cite{Meerburg:2015zua}.
\be
N_{eff}=3.046+\left(3.046+\frac{8}{7}\left(\frac{11}{4}\right)^{4/3}\right)\frac{1}{12}\int d \ln k \, P_T \label{eq:neff}
\ee
where $N_{eff}=3.046$ is the Standard Model prediction and current $68\%$ confidence level suggest $\Delta N_{eff}\leq0.19$. Thus, $N_{eff}$ provides an indirect probe of all scales of the GW spectrum. It is important to note that as long as $r$ is not measured, \eqref{eq:neff} holds limited promise, as the amplitude $A_T$ and therefore $r$ can be arbitrarily small. Bouncing models, for instance, predict $r<10^{-30}$, and only a handful suggest an observable $r$ \cite{Ben-Dayan:2016iks,Ben-Dayan:2018ksd}.

A similar situation of inaccessibility occurs for the scalar spectrum, where we have so far probed only $8$ out of the expected $50-60$ 'e-folds' of inflation. This limit is not expected to improve in the near future, due to built-in non-linearities.
Nevertheless, indirect probes have provided useful indications and constraints on the scalar spectrum  \cite{Chluba:2012we,Ben-Dayan:2013eza,Ben-Dayan:2014iya,Ben-Dayan:2015zha,Bringmann:2011ut,Vieira:2017oqg} on scales beyond primary CMB scales.

The above discussion implicitly assumed full decoupling between the scalar and tensor modes. It is valid in first order in perturbation theory. However, at second order, scalar fluctuations act as sources of tensor fluctuations \cite{Mollerach:2003nq, Ananda:2006af, Baumann:2007zm}.
These induced, second order tensor fluctuations are related to the scalar (first order) spectrum via $P_T^{(2)}\sim P_S^2$. Given the already measured scalar spectrum \eqref{eq:PSplanck}, one is guaranteed a tensor signal at the level of $r\sim 10^{-6}$ on CMB scales. If, in the distant future, such a signal is not measured, we have misinterpreted our Early Universe paradigm or detected violations of  general relativity. Analysis related to this second order GW spectrum and its phenomenological consequences has recently been discussed in \cite{Alabidi:2012ex,Kohri:2018awv,Byrnes:2018txb,Inomata:2018epa}

Given this induced GW spectrum, it can be constrained or measured by LI and PTA experiments. Furthermore, by adapting \eqref{eq:neff} to the induced spectrum, it will also contribute to $N_{eff}$. Due to its functional dependence on the scalar spectrum, LI and PTA provide an indirect probe of the scalar spectrum on relevant scales. Better yet, $N_{eff}$ will now be sensitive to \textit{all scales} of the \textit{scalar spectrum}. Hence, we have novel probes of the scalar power spectrum at scales inaccessible to primary CMB constraints. Moreover this indirect $N_{eff}$ constraint is based on CMB data alone.

In this paper, we analyze how $N_{eff}$, LIGO and PTA data constrain the various parameterizations of the scalar power spectrum and give forecasts for future experiments that are the Simons Observatory (SO), Stage 4, SKA-PTA and LISA.
Given that $N_{eff}$ will include an integral over all scales it will be sensitive to enhancements of the spectrum and to the smallest scales, i.e. the UV cut-off.
While each parameterization has limitations, it still uses a minimal number of parameters and avoiding our conclusions require additional parameters that make the spectrum more fine-tuned and less plausible. We find that $N_{eff}$ strongly constrains deviations from the constant $n_s$ spectrum. In particular it forces the running and the running-of-running to conform to the standard slow-roll hierarchy, $\alpha,\beta \lesssim (n_s-1)^2$. Such a constraint rules out a large portion of the parameter space allowed by Planck. We further perform a likelihood analysis including the $N_{eff}$ constraint and find that positive $\alpha, \beta$ are disfavored by at least $1\sigma$ and  $\beta + 0.074\ \alpha>8.6\times 10^{-4}$ are ruled out at $3\sigma$. These constraints rule out certain scenarios for primordial black holes formation (PBH), that have recently emerged as possible dark matter candidates \cite{Munoz:2016owz}. 

The paper is organized as follows, in section \ref{sec:Relevant} we list the different experiments we are interested in and their forecasts for relevant parameters. In section \ref{sec:2t} we reproduce the major steps leading to the induced tensor spectrum, and the parameterizations of the scalar spectrum we are interested in. In section \ref{sec:analytic} we calculate the predicted $N_{eff}$ and the allowed parameter space. In section \ref{sec:Likelihood} we report the results of a likelihood analysis. 
In section \ref{sec:constraints} we discuss constraints on the spectrum due to present day measurements. We then conclude.

\section{Relevant Experiments}
\label{sec:Relevant}
We divide the different experiments into two categories. One category is the CMB observations that probe the GW energy density at the time of decoupling. Usually it is specified in terms of $r$, the tensor-to scalar ratio and wavenumber $k$. These measurements probe the GW energy density on largest scales, $k\lesssim k_{eq}$. Furthermore, CMB observations also probe $N_{eff}$.  The second category probes the GW energy density today such as LIGO and PTA. In these cases, one usually writes down the fractional energy density, $\Omega_{GW}$ as a function of frequency $f$. These measurements probe much smaller scales with $k \gg k_{eq}$.
In Table \ref{table:CMB} we list the different experiments and their forecasted sensitivities. The table allows simple comparison between the different scales and sensitivities. The conversion between frequencies and wavenumbers is 
\be \label{eq:ftok}
f=\frac{k c}{2\pi a(\eta_0)}=1.5\times 10^{-15}\, \left(\frac{k}{ \mathrm{ Mpc^{-1}}}\right)\, \mathrm{Hz},
\ee
where we have used $c=3\times 10^8\, m/s$ as the speed of light.
The relation between $P_T$ and present day $\Omega_{GW}$ is \cite{Baumann:2007zm}:
\be
\Omega_{GW}=4.2\times10^{-2}r A_s\left(\frac{k}{k_0}\right)^{n_T}\frac{a_{eq}}{a(\eta_0)}\; .
\ee

Carrying out the CMB experiments up to Stage 4 will decisively constrain a narrow space of allowed inflationary models, or will rule out all large field models. Besides specifying the forecasts for various experiments, we performed a Fisher matrix analysis of cosmic variance limited (CVL) CMB polarization measurement.  
The analysis predicts $\sigma(r)=2.2 \times 10^{-6}$ for $f_{sky}=0.8$ with no delensing and ignoring beam systematics. Such value seems very close to the induced second order GW spectrum prediction of $r\sim 10^{-6}$, which is close to a guaranteed signal.
\begin{table}[!h]
 
    \begin{tabular}{||c|c|c|c|c|c||}
        \hline
         Experimental stage &$r<$&$\Delta N_{eff}$ &$\Omega_{GW}<$ & Wavenumber $\mathrm{Mpc^{-1}}$ & frequency $\mathrm{Hz}$ \\
         \hline
        present & $0.06$ & $0.19$ & $1.6 \times 10^{-15}$ & $0.05$ & $7.5 \times 10^{-17}$  \\
         S2 & $0.035$ & $0.14$ & $9.1 \times 10^{-16}$ & $0.05$  & $7.5 \times 10^{-17}$ \\
         SO & 0.006 &0.04 &  $1.6 \times 10^{-16}$ & 0.05 & $7.5 \times 10^{-17}$  \\
         S4& 0.0005 &0.027 & $1.3 \times 10^{-17}$ &0.05 & $7.5 \times 10^{-17}$  \\
         CVL & $2.2 \times 10^{-6}$ & $3.1 \times 10^{-6}$ & $5.6 \times 10^{-20}$ &0.05 & $7.5 \times 10^{-17}$  \\
         \hline
        LIGO & $6.5 \times 10^6$ & NA & $1.7\times 10^{-7}$ & $(3-13)\times 10^{16}$ &20-86 \\
         aLIGO & $3.86\times 10^4$ & NA &  $10^{-9}$ & $(3-13)\times 10^{16}$ &$\sim 50$\\
          PTA &$5.01 \times 10^4$ & NA & $1.3 \times 10^{-9}$ & $1.5\times 10^8$ & $10^{-7}$\\
         SKA-PTA & 50.1 &NA & $1.3 \times 10^{-12}$ & $1.5\times 10^8$ & $10^{-7}$\\
         LISA  & 3.86 & NA & $10^{-13}$ & $(1.5-15) \times 10^{12}$ &  0.001-0.01\\
         \hline
    \end{tabular}
    \caption{The forecast of constraints on $r$ and $\Delta N_{eff}$ for different experiments. The details were taken from \cite{Ade:2018sbj,Abazajian:2016yjj}. The quoted bound on $r$ for the CMB future probes is the forecast for $\sigma(r)$. The CVL result is based on a Fisher matrix analysis.} \label{table:CMB}
\end{table}

\section{Gravitational waves spectrum induced by primordial scalar perturbations}
\label{sec:2t}

 In this section we repeat the main steps in the calculations done in \cite{Mollerach:2003nq,Ananda:2006af,Baumann:2007zm} and derive an expression for the contribution of the induced GW  to $N_{eff}$.
The bottom line is that we have an expression for the GW energy density for all times.
We start from the background Friedmann equations:
\beq
\cH^2 = \frac{\kappa^2 a^2}{3} \rho^{(0)}\, ,
\quad \cH^2- \cH'  = \frac{\kappa^2 a^2}{2} (\rho^{(0)} + P^{(0)})\, ,
\quad \cH \equiv \partial_\eta \ln a\, .
\eeq
Here $\rho^{(0)}$ and $P^{(0)}$ are the homogeneous background density and pressure, respectively, and the prime denotes a derivative with respect to conformal time, $\eta$.
At linear order in perturbation theory, different $k$-modes in Fourier space are independent. In the absence of an external source, the mode equation for the tensor perturbation reads, $Q_k=a h_k$ :
\be
Q_k''+\left(k^2-\frac{a''}{a}\right)Q_k=0 \; .
\ee
This is in contrast to the second order Einstein equations, $G^{(2)}_{\mu \nu} = \kappa^2 T^{(2)}_{\mu \nu}$, where different $k$-modes mix and scalar, vector and tensor modes are not independent. 
Instead, there will be a second-order contribution to the tensor mode, $h_{ij}^{(2)}$, that depends quadratically on the first-order scalar metric perturbation.  

Consider the FLRW metric perturbed up to second order,
\beq
\label{equ:metric}
\d s^2 = a^2(\eta) \left[-\left(1 + 2 \Phi^{(1)} + 2 \Phi^{(2)} \right) \d 
\eta^2 + 2 V^{(2)}_i \d \eta \d x^i
       + \left\{ \left(1 - 2 \Psi^{(1)} - 2 \Psi^{(2)} \right) \delta_{ij} + 
\frac{1}{2} h_{ij} \right\} \d x^i \d x^j \right]\, ,
\eeq
where 
$h_{ij} \equiv h^{(2)}_{ij}$ and we have ignored first-order vector and tensor 
perturbations. 
The projected Einstein equations with the tensor $\hat{\cal T}_{ij}^{~lm}$ are given by \cite{Ananda:2006af},
\beq
\label{equ:E2}
\hat{\cal T}_{ij}^{~lm} G^{(2)}_{lm} = \kappa^2 \hat{\cal T}_{ij}^{~lm} 
T^{(2)}_{lm}\, .
\eeq
and the mode equation gets a source term: 
\beq
\label{equ:hh}
h''_{ij} + 2 \cH h'_{ij} - \nabla^2 h_{ij}
= - 4 \hat{\cal T}_{ij}^{~lm} {\cal S}_{lm}\, ,
\eeq
with
\bea
{\cal S}_{ij}
\equiv 2 \Phi \partial^i \partial_j \Phi - 2 \Psi \partial^i \partial_j \Phi
    + 4 \Psi \partial^i \partial_j \Psi
     +& \partial^i \Phi \partial_j \Phi - \partial^i \Phi \partial_j \Psi
     - \partial^i \Psi \partial_j \Phi + 3 \partial^i \Psi \partial_j \Psi &\\ \nonumber
- \frac{4}{3(1+w) \cH^2} \partial_i \left( \Psi' + \cH \Phi \right)
      \partial_j \left( \Psi' + \cH \Phi \right) 
   &  -\frac{2 c_{s}^{2}}{3 w \cH^2} \left[ 3 \cH (\cH \Phi - \Psi') + \nabla^2 
\Psi \right]
      \partial_i \partial_j (\Phi - \Psi)\, .\cr &
\eea
where, $w \equiv P^{(0)}/{\rho^{(0)}}$, is the equation of state parameter, $\Phi \equiv \Phi^{(1)}$ and $\Psi \equiv 
\Psi^{(1)}$.
The Fourier transform of tensor metric perturbations is
\beq
h_{ij}(\mathbf{x},\eta) = \int \frac{\d^3 \mathbf{k}}{(2\pi)^{3/2}} e^{i 
\mathbf{k} \cdot \mathbf{x}} \left[ h_{\mathbf{k}}(\eta) 
\mathsf{e}_{ij}(\mathbf{k}) + \bar h_{\mathbf{k}}(\eta) \mathsf{\bar 
e}_{ij}(\mathbf{k}) \right]\, ,
\eeq
where the two time-independent polarization tensors $\mathsf{e}_{ij},\mathsf{\bar e}_{ij}$ are written in terms of the orthonormal basis vectors 
$\mathbf{e}$ and $ \bar{\mathbf{e}}$ orthogonal to $\mathbf{k}$,
\bea
\mathsf{e}_{ij}(\mathbf{k}) &\equiv& \frac{1}{\sqrt{2}} 
[\mathsf{e}_i(\mathbf{k}) \mathsf{e}_j(\mathbf{k}) - \mathsf{\bar 
e}_i(\mathbf{k}) \mathsf{\bar e}_j(\mathbf{k})]\, ,\\
\mathsf{\bar e}_{ij}(\mathbf{k}) &\equiv& \frac{1}{\sqrt{2}} 
[\mathsf{e}_i(\mathbf{k}) \mathsf{\bar e}_j(\mathbf{k}) + \mathsf{\bar 
e}_i(\mathbf{k}) \mathsf{e}_j(\mathbf{k})]\, .
\eea
The equation of motion for the gravitational wave amplitude for both $h$
and $\bar h$ reads
\beq
h''_{\mathbf{k}} + 2 \cH h'_{\mathbf{k}} + k^2 h_{\mathbf{k}}=  {\cal 
S}(\mathbf{k},\eta)\, ,
\label{EOMh}
\eeq
where the source term, ${\cal S}$, is a convolution of two first-order scalar perturbations,
\bea
\label{equ:source}
{\cal S}(\mathbf{k},\eta)
&=& -4 \mathsf{e}^{lm}(\mathbf{k} ) {\cal S}_{lm}(\mathbf{k})\cr
&=& 4 \int \frac{\d^3 \mathbf{\tilde k}}{(2\pi)^{3/2}} 
\mathsf{e}^{lm}(\mathbf{k}) {\tilde k}_l {\tilde k}_m
    \Biggl[ \left\{ \frac{7+3w}{3(1+w)} - \frac{2 c_s^2}{w} \right\}
            \Phi_{\mathbf{\tilde k}}(\eta) \Phi_{\mathbf{k-\tilde k}}(\eta)
            + \left( 1 - \frac{2 c_s^2 {\tilde k}^2}{3 w \cH^2} \right)
              \Psi_{\mathbf{\tilde k}}(\eta) \Psi_{\mathbf{k-\tilde k}}(\eta) 
\nonumber \\
& &         + \frac{2 c_s^2}{w} \left( 1 + \frac{{\tilde k}^2}{3 \cH^2} \right)
              \Phi_{\mathbf{\tilde k}}(\eta) \Psi_{\mathbf{k-\tilde k}}(\eta)
            + \left\{ \frac{8}{3(1+w)} + \frac{2 c_s^2}{w} \right\} 
\frac{1}{\cH}
              \Phi_{\mathbf{\tilde k}}(\eta) \Psi'_{\mathbf{k-\tilde k}}(\eta) 
\nonumber \\
& &         - \frac{2 c_s^2}{w \cH}
              \Psi_{\mathbf{\tilde k}}(\eta) \Psi'_{\mathbf{k-\tilde k}}(\eta)
            + \frac{4}{3(1+w) \cH^2}
              \Psi'_{\mathbf{\tilde k}}(\eta) \Psi'_{\mathbf{k-\tilde k}}(\eta)
    \Biggr]\, . \label{equ:Source2}
\eea

The particular solution of (\ref{EOMh}) is then derived using the Green's 
function:
\beq
\label{equ:solh}
h_\vk(\eta) =  \frac{1}{a(\eta)} \int \d\tilde \eta \,  g_\vk(\eta;\tilde \eta) 
\Bigl[ a(\tilde \eta) {\cal S}(\vk,\tilde \eta) \Bigr]\, ,
\eeq
where 
\beq
\label{equ:g}
g''_\vk + \Bigl( k^2 - \frac{a''}{a} \Bigr) g_\vk  = \delta(\eta-\tilde\eta)\, .
\eeq

Evaluating the two-point correlation function results in
\beq
\langle h_{\mathbf{k}}(\eta) h_{\mathbf{K}}(\eta) \rangle
= \frac{1}{a^2(\eta)} \int_{\eta_0}^\eta \d \tilde \eta_2 \int_{\eta_0}^\eta \d 
\tilde \eta_1 \,
  a(\tilde \eta_1) a(\tilde \eta_2) g_\vk(\eta; \tilde \eta_1) g_{\vK}(\eta; \tilde 
\eta_2)\,
  \langle {\cal S}(\mathbf{k}, \tilde \eta_1) {\cal S}(\mathbf{K}, \tilde 
\eta_2) \rangle\, ,
\label{h-correlation}
\eeq
and its relation to the power spectrum, is defined as:
\beq
\label{equ:Ph}
\langle h_{\mathbf{k}}(\eta) h_{\mathbf{K}}(\eta) \rangle = \frac{2 \pi^2}{k^3} 
\delta(\mathbf{k} + \mathbf{K}) P_T(k, \eta)\, .
\eeq


Hence, the fractional energy density of the induced GW is then given by
\be
\Omega^{(2)}_{GW}(k, \eta)=\frac{k^2}{6\pi^2 \cH^2}t^2(k,\eta)P^{\rm (2)}_T(k)=\frac{a(\eta)k^2}{a_{eq}k_{eq}^2} t^2(k,\eta)P^{\rm (2)}_T(k)\; .
\ee
The power spectrum at horizon crossing therefore scales as follows
\beq
\label{equ:FinalPh}
P^{\rm (2)}_T(k) \propto P_S^2 \left\{ \begin{array}{cc} 
\frac{k_{\rm eq}}{k} & \ \quad k < k_{\rm eq} \\  1 & \ \quad  k > k_{\rm eq} 
\end{array} \right. \; .
\eeq
The transfer function $t(k,\eta)$ is approximated by
\beq
\label{equ:Finalt}
t(k, \eta) = \left\{ \begin{array}{l l} 1 & \ \quad k < k_{\rm eq} \\  \Bigl( 
\frac{k}{k_{\rm eq}}\Bigr)^{-\gamma} & \ \quad k_{\rm eq} < k < k_c(\eta) \\  
\frac{a_{\rm eq}}{a(\eta)} \frac{k_{\rm eq}}{k} & \ \quad  k > k_c(\eta) 
\end{array} \right.\; .
\eeq
Inserting the transfer function and the induced spectrum gives:
\be
\label{eq:o2gw}
\Omega^{(2)}_{GW}(k, \eta)=A_{GW}^{(2)}P_S^2(k)f(k, k_{eq},a(\eta),a_{eq})
\ee
\be
\label{eq:fo2gw}
f=\left\{ \begin{array} {cc} \frac{a(\eta)}{a_{eq}}\frac{k}{k_{eq}} & \ \quad k<k_{eq}\\
 \frac{a(\eta)}{a_{eq}}\left(\frac{k}{k_{eq}}\right)^{2-2\gamma} & \ \quad k_{eq}<k<k_c(\eta)\\
\frac{a_{eq}}{a(\eta)} & \ \quad k>k_c(\eta) 
\end{array} \right.\; .
\ee
Numerically it turns out $\gamma \simeq 3,A_{GW}^{(2)}\simeq 10$. The behavior of large $k$ is due to subhorizon modes that have not settled down yet.
It is given by:
\be
k_c(\eta)=\left(\frac{a(\eta)}{a_{eq}}\right)^{1/(\gamma-1)}k_{eq}
\ee
The modes that have not settled down will be our primary interest for two reasons. First, considering CMB observations and limits on $N_{eff}$, part of $N_{eff}$ is the integral over all wavenumbers of the stochastic GW background. If the GW spectrum is blue, the unsettled modes will be a dominant contribution to $N_{eff}$.  Second, the GW spectrum today, is directly probed by LI and PTA experiments. These experiments only probe a limited domain of scales. These scales again are the ones related to modes that have not settled down. 

Punching in the numbers, for CMB we have $a_{CMB}\simeq 3a_{eq}$, the relevant $k_c(\eta_{CMB})\simeq \sqrt{3} k_{eq}$ and $f(k>\sqrt{3} k_{eq})=1/3$.
Hence $k>\sqrt{3} k_{eq}$ are the modes of interest when we wish to determine the effect of second order GW on $N_{eff}$.
For the LI and PTA experiments, 
$a_{today}\equiv a(\eta_0)\simeq 3400 a_{eq}$, yielding $k_c(\eta_0)\simeq 58 k_{eq} \sim 1 \mathrm{Mpc^{-1}}$. We shall use it to put direct constraints on the  primordial scalar power spectrum using LIGO and PTA measurements and provide a forecast for future experiments.

The dependence of $N_{eff}$ on the GW spectrum is given by \cite{Meerburg:2015zua}:
\be
\label{eq:pt}
N_{eff}=3.046+\left(3.046+\frac{8}{7}\left(\frac{11}{4}\right)^{4/3}\right)\frac{1}{12}\int d \ln k \, P_T \; .
\ee
Substituting $P_T(k>k_{eq})=A_{GW}^{(2)}P_S^2/3$ gives
\be
\label{eq:ps2}
N_{eff}=3.046+\left(3.046+\frac{8}{7}\left(\frac{11}{4}\right)^{4/3}\right)\frac{A_{GW}^{(2)}}{36}\int^{k_{UV}} d \ln k \, P_S(k)^2\; .
\ee
Using the Planck data \cite{Akrami:2018odb,Ade:2018gkx}, 
we have a bound at $95\%$ confidence level of $2\Delta N_{eff}\leq 0.38$, therefore:
\be
\label{eq:bound}
I\equiv \int^{k_{UV}} d \ln k \, P_S(k)^2 \leq 0.18\; ,
\ee
while for later CMB experiments right hand side of the bound will improve to $I<0.14$ for S2, $I<0.05$ for the Simons Observatory \cite{Ade:2018sbj}, and $I<0.03$ for Stage 4. Notice that $N_{eff}$ allows us to probe indirectly \textit{all scales} of both the scalar and GW power spectrum.
By considering various forms of the scalar spectrum we can constrain its parameters measuring $N_{eff}$. This is one of the major results of this work. 
For this purpose, we use several common parameterizations of the spectrum:

\begin{align}
P_S &=A_s\left(\frac{k}{k_0}\right)^{n_s(k_0)-1}, \quad & (const.)\\
P_S &=A_s\left(\frac{k}{k_0}\right)^{n_s(k_0)-1+\frac{\alpha(k_0)}{2} \ln\frac{k}{k_0}+\frac{\beta(k_0)}{6} \ln^2\frac{k}{k_0}}\,, \quad & (run) \label{eq:run}\\
P_S&=A_s \left(\frac{k}{k_0}\right)^{n_s(k_0) - 1} + B \left(\frac{\pi e}{3}\right)^{3/2} \left(\frac{k}{k_i}\right)^3 e^{-\pi/2 (k/k_i)^2}\,, \quad  & (bump)\label{eq:bump} \\
P_S&=A_s \left(\frac{k}{k_0}\right)^{n_s(k_0) - 1} \left[1+\frac{B}{A_s}\Theta(k-k_i)\right]\,, \quad & (step) \label{eq:step}\\
P_S&=A_s \left(\frac{k}{k_0}\right)^{n_s(k_0) - 1} \left[\Theta (k_i-k)+\left(\frac{k}{k_i}\right)^{n_s^*(k_0)-1}\Theta(k-k_i)\right]. \quad & (bend) \label{eq:bend}
 \end{align}

Throughout this work we use Planck's maximal likelihood value of $A_s=2.1\times 10^{-9}$. Here $\alpha$ is dubbed the 'running of the spectral index' and $\beta$ the 'running of running'. A word of caution is that usually the parameterizations above are limited to CMB scales while extrapolating them to all scales of inflation might be problematic. However, these parameterizations are suitable for a wide variety of models and further modifications mean that the spectrum and the underlying inflationary model are more complicated and not generic. Further complications or features of the spectrum can always be found. In such a case, our main constraint \eqref{eq:bound}, can be calculated and used to test the spectrum. 

\section{Constraining the scalar spectrum using $N_{eff}$}
\label{sec:analytic}
To a good approximation the integral in \eqref{eq:bound} can be approximated as $\int^{k_{UV}} d \ln k P_S(k)^2\simeq  P_S(k_{UV})^2/2(n_s(k)-1)$.
It will therefore be very sensitive to continuous rise in power like with running or a bend, but will be rather insensitive to features localized in $k$ or an increase in the amplitude but not in the tilt (a step). Due to the sensitivity to $k_{UV}$, we shall consider several scenarios, $k_{UV}/k_0=10^{21},\,10^{24},\,10^{27},\,10^{30}$. The first corresponds to two decades beyond LIGO scales and is not theoretically motivated, the second to $60$ e-folds of inflation, the third to $67$ e-folds and the last to the Planck scale. In all the following figures, shaded regions mean they are excluded regions in violation of the $N_{eff}$ measurement by $2\sigma$ or more. The bounds on $\Delta N_{eff}$ are given in table \ref{tab:S_i}:
\begin{table}[!h]
    \centering
    \begin{tabular}{||c|c||}
        \hline
         Stage&$S_i=2\Delta N_{eff}$\\
         \hline
         current&0.38  \\
         S2&0.28  \\
         SO&0.11  \\
         S4&0.054 \\
         \hline
    \end{tabular}
    \caption{The forecast of constraints on $2\Delta N_{eff}$ for different CMB experiment stages (taken from \cite{Ade:2018gkx,Akrami:2018odb,Ade:2018sbj,Abazajian:2016yjj}).}
    \label{tab:S_i}
\end{table}

\begin{figure}[!ht]
\begin{center}
\includegraphics[width=0.75\textwidth]{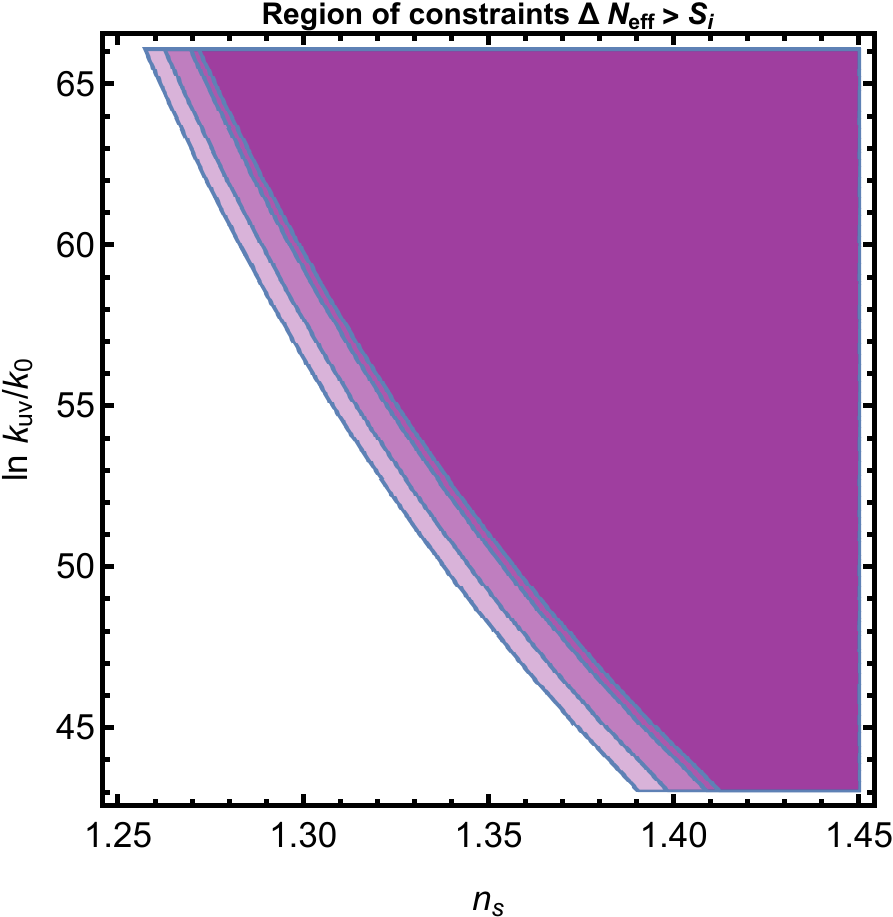}
\caption{Although a purely academic exercise, nevertheless it is interesting to see how $N_{eff}$ constrains the blue scalar index $n_s$ values. The different shades of purple correspond to different cosmology experiments generations, where the most restrictive is S4, and the least is current data. The vertical axis is the UV scale, which is the upper limit for our integration over $P_S$, to yield the $\Delta N_{eff}$ as allowed by MCMC analyses and other inputs.}
\label{fig:ns_kuv_ira}
\end{center}
\end{figure}
\subsection{$n_s=const.$}
The simplest case we start with is a constant $n_s$. Since we know that $n_s\simeq0.97$, this is a purely academic exercise. Nevertheless, substituting a constant $n_s>1$ yields
\be
I\equiv \int^{k_{UV}} d\ln k \, P_S^2=\frac{A_s^2}{2(n_s-1)}\left(\frac{k_{UV}}{k_0}\right)^{2(n_s-1)},
\ee
and the resulting constraints are given in Fig. \ref{fig:ns_kuv_ira}. Selected results are also given in the following Table \ref{tab:ns}.
\begin{table}[!h]
    \centering
    \begin{tabular}{||c|c|c||}
         \hline
         \hline
         Stage&$\mathrm{\log_{10}[k_{UV}/k_0]}$ &Constrained by $n_s< $\\
         \hline
         \hline
         Current&\begin{tabular}{c}
              21\\
              24\\
              27\\
              30
         \end{tabular}
              & \begin{tabular}{c}
              1.412\\
              1.347\\
              1.304\\
             1.273
         \end{tabular}\\
         \hline
         S4&\begin{tabular}{c}
              21\\
              24\\
              27\\
              30
         \end{tabular} & \begin{tabular}{c}
        1.391\\
        1.326\\
        1.286\\
        1.257
         \end{tabular}\\
         \hline
         \hline
    \end{tabular}
    \caption{Constraints on a constant the spectral index $n_s$ as given by different UV cutoffs for current data and forecasts for S4.}
     \label{tab:ns}
\end{table}

\subsection{Running spectral index}
Next we consider the case of running and running of running, \eqref{eq:run}. Such parameterizations provide a good fit to the low multipole power deficit \cite{Ade:2015xua,Akrami:2018odb,Ade:2018gkx}. 
In the case of vanishing $\beta$, there is a simple analytic expression for the integral $I$:
\be
I\equiv \int^{k_{UV}} d\ln k \, P_s^2(\alpha\neq 0,\beta=0)=\frac{\sqrt{\pi } }{2 \sqrt{\alpha }}A_s^2e^{-\frac{(n_s-1)^2}{\alpha }} \text{erfi}\left(\frac{n_s-1+\alpha  \ln \left(\frac{k_{UV}}{k_0}\right)}{\sqrt{\alpha }}\right)\;.
\ee
For $\beta\neq 0$ there is no simple analytical expression.
\begin{table}[!h]
    \centering
    \begin{tabular}{||c|c|c||}
         \hline
         \hline
         Stage&$\log_{10}[k_{UV}/k_0]$ &$\alpha$ Constrained by (at $n_s=0.97)$ \\
         \hline
         \hline
         Current&\begin{tabular}{c}
              21\\
              24\\
              27\\
              30
         \end{tabular}
              & \begin{tabular}{c}
              0.0178\\
              0.0137\\
              0.0109\\
              0.0089
         \end{tabular}\\
         \hline
         S4&\begin{tabular}{c}
              21\\
              24\\
              27\\
              30
         \end{tabular} & \begin{tabular}{c}
        0.017\\
        0.0131\\
        0.0105\\
        0.00855
         \end{tabular}\\
         \hline
         \hline
    \end{tabular}
    \caption{Constraints on the running of the spectral index in case of a fixed running ($\beta=0$), as given by different UV cutoffs, in the current stage as well as in S4.}
    \label{tab:running_index}
\end{table}

The results for current data and the forecast for S4 experiment for varying $k_{UV}/k_0$ is given as a function of $\alpha$ with $n_s=0.97,\beta=0$  in Table \ref{tab:running_index}. 
\begin{figure}[!h]
\begin{center}
\includegraphics[width=0.75\textwidth]{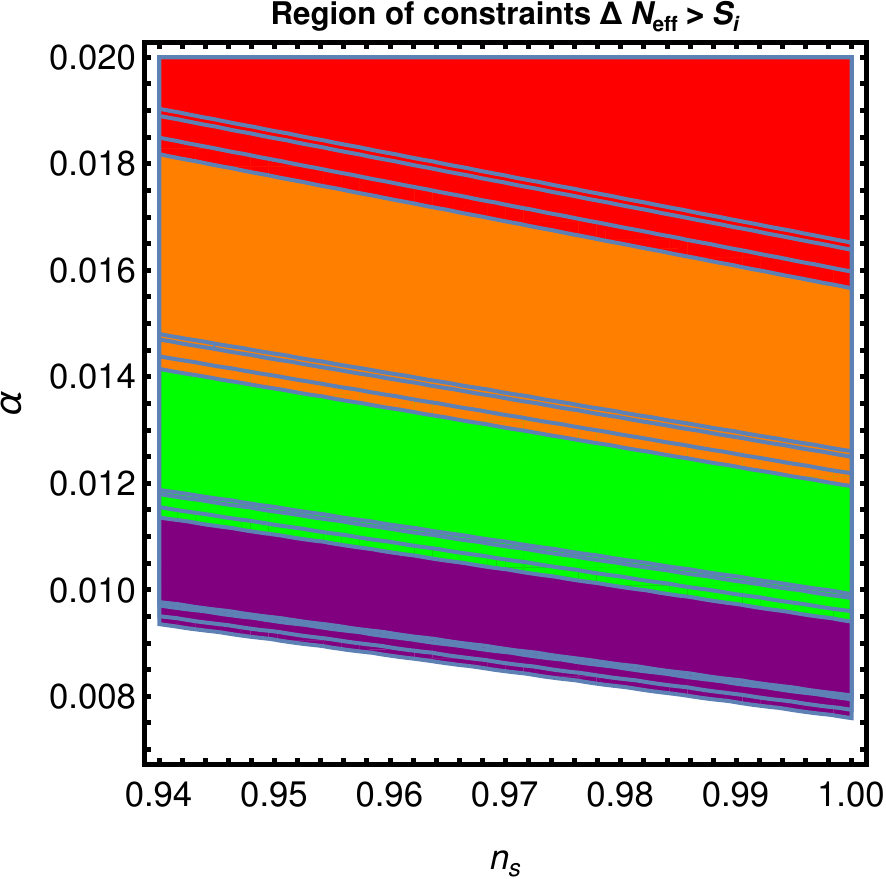}
\caption{The constraints on such a power spectrum are shown above where the purple region corresponds to $k_{UV}/k_0=10^{30}$, orange to $k_{UV}/k_0=10^{27}$, green to $k_{UV}/k_0=10^{24}$ and red corresponds to $k_{UV}/k_0=10^{21}$. Each internal line is given by restrictions on $\Delta N_{eff}$. These are given by: $S_i = (0.38,0.28,0.11,0.054)$, for the different observational stages.
\label{fig:ns_alpha_kuv}}
\end{center}
\end{figure}
The full results are given in Fig. \ref{fig:ns_alpha_kuv} for various UV scales and spectral index $n_s$. 

\begin{figure}[!h]
\begin{center}
\includegraphics[width=0.75\textwidth]{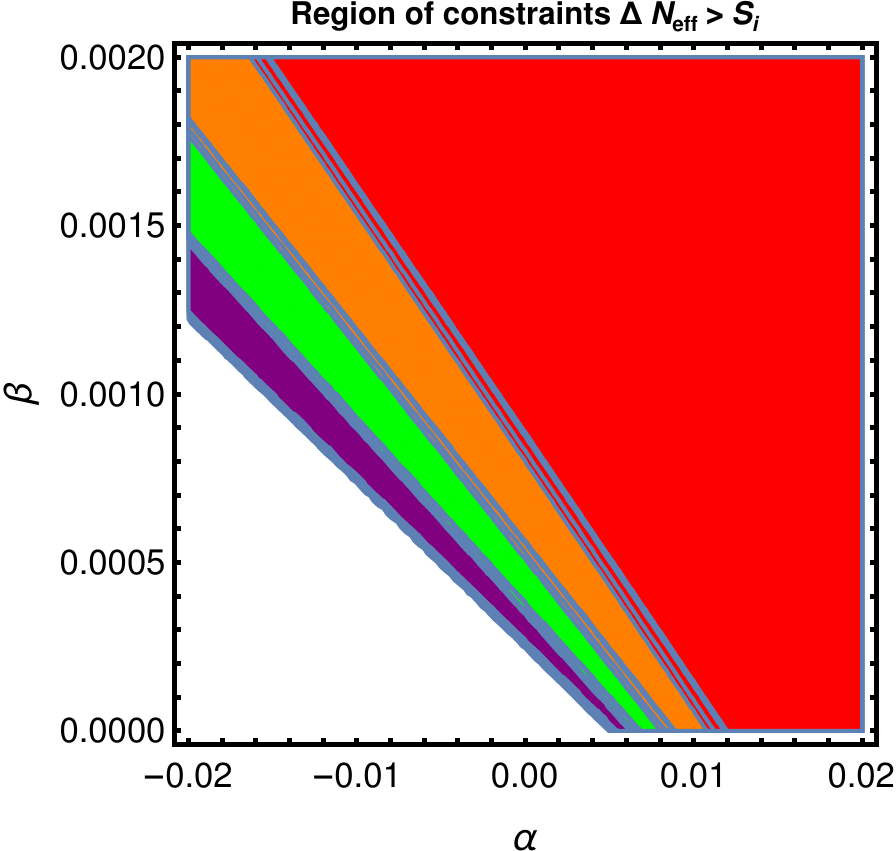}
\caption{An analysis of constraints for the case of running  spectral index ($\alpha$) with running of running ($\beta$). Each color corresponds to a different UV scale cutoff, purple for $k_{UV}/k_0 = 10^{30}$, to $10^{27}$, $10^{24}$ down to red for $k_{UV}/k_0 = 10^{21}$. The different same color lines correspond to the different CMB experimental stages $S_1-S_4$.}
\label{fig:alpha_beta_ira}
\end{center}
\end{figure}

Next we consider the inclusion of "running of running", $\beta$. 
It is further constrained giving a conservative bound with current data, as presented in Table \ref{tab:alpha_beta}.
\begin{table}[!h]
    \centering
    \begin{tabular}{||c|c|c||}
         \hline
         \hline
         Stage&$\mathrm{\log_{10}[k_{UV}/k_0]}$ &$\beta$ Constrained by (at $\alpha=0, n_s=0.97$)   $\times 10^{-4}$ \\
         \hline
         \hline
         Current&\begin{tabular}{c}
              21\\
              24\\
              27\\
              30
         \end{tabular}
              & \begin{tabular}{c}
              11.15\\
              7.53\\
              5.32\\
              3.91
         \end{tabular}\\
         \hline
         S4&\begin{tabular}{c}
              21\\
              24\\
              27\\
              30
         \end{tabular} & \begin{tabular}{c}
        10.7\\
        7.2\\
        5.1\\
       3.75
         \end{tabular}\\
         \hline
         \hline
    \end{tabular}
    \caption{Constrains on index running of running ($\beta$) at a vanishing index running ($\alpha=0$), for current stage and Stage 4 cosmology. }
    \label{tab:alpha_beta}
\end{table}
 The full results of this study are given in Fig. \ref{fig:alpha_beta_ira} where both $\alpha$ and $\beta$ are allowed to be positive.
 It is evident that a case of positive $\beta$ is heavily constrained to be of the order of $(n_s -1)^2$ for the values of $\alpha$ currently allowed by CMB analysis.

\begin{figure}
\centerline{\includegraphics[width=0.9\textwidth]{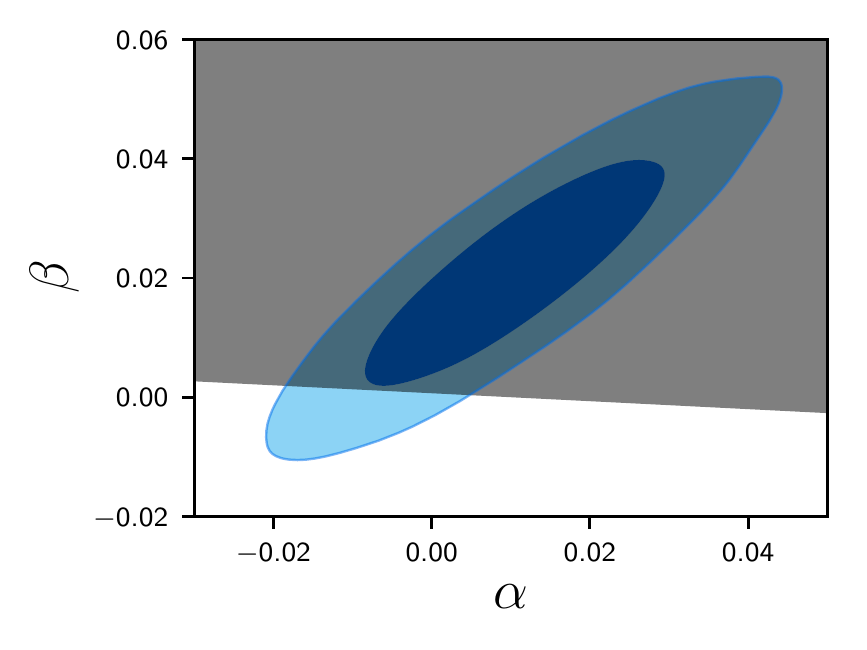}}
\caption{Planck 2015 constraints on running $\alpha$ and running of running $\beta$. Taking into account the contribution of 2nd order tensors to $N_{eff}$, the shaded region is ruled out assuming $k_{UV}/k_0=10^{21}$ and $n_s=0.97$.}
\label{fig:shaded_likelihood}
\end{figure}

Finally, the discriminatory power of the method, is best presented by overlaying the results on top of likelihood contours of existing data. This is depicted in Fig. \ref{fig:shaded_likelihood}, taken from Planck 2015 data with $n_s=0.97, k_{UV}/k_0=10^{21}$. 
Thus, slow-roll hierarchy must be maintained, and in particular having either $\alpha$ or $\beta\sim (n_s-1)$ will violate the $N_{eff}$ bound.
So taking this result at face value means that such runnings cannot explain the low multipoles power deficit.
In \cite{Munoz:2016owz}, the requirement was $\beta>0.03$ for PBHs to be a DM candidate, and $\beta\gtrsim 0.002$ to produce PBH with mass $M>10^{15}\,gr$. Here we show that both are in violation of the bound from $N_{eff}$. 
One can further consider higher and higher orders of scale dependence, such as $\gamma= d^3 n_s/d \ln k^3$. As our analysis shows, any such higher order term will be constrained more severely, and may not help as well.
The only way for PBH to be DM is if they were serendipitously produced at some narrow range of wavenumbers, as the spectrum with a bump suggests. 
Hence, our result confirms that the only valid models are the ones where the slow-roll hierarchy is maintained and at most $\alpha,\beta\lesssim few \times (n_s-1)^2$ for positive running.

\subsection{Spectrum with a bump/particle production}
\begin{figure}[!h]
\begin{center}
\includegraphics[width=0.5\textwidth]{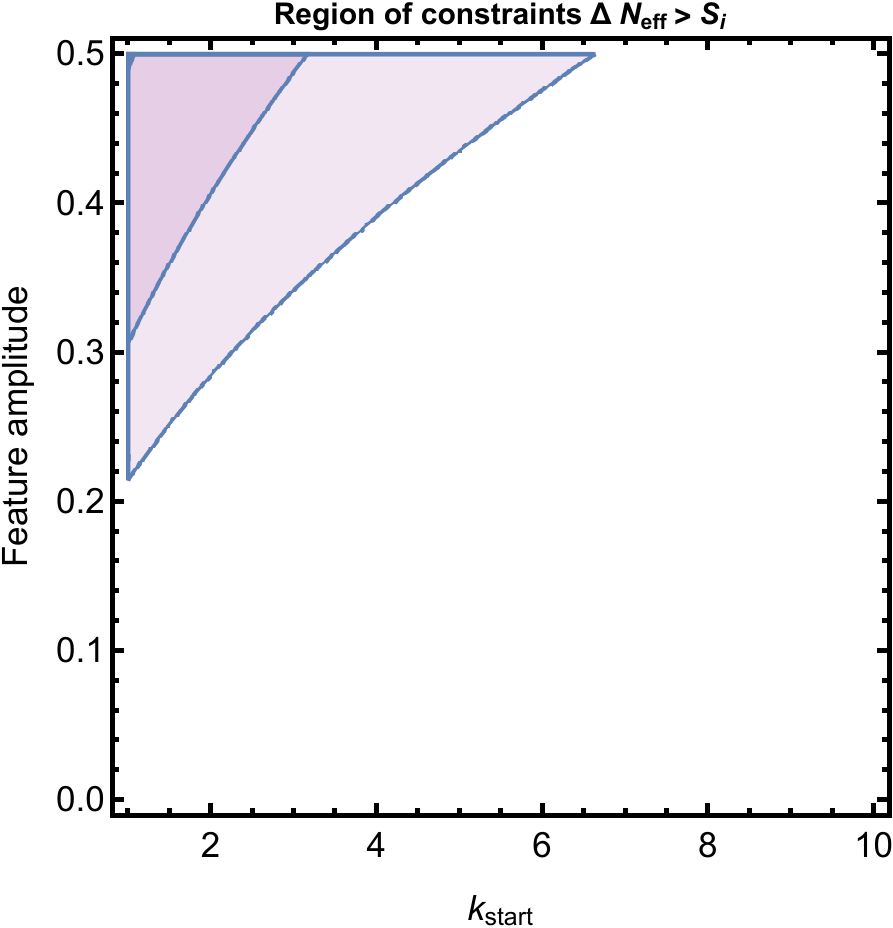}\includegraphics[width=0.5\textwidth]{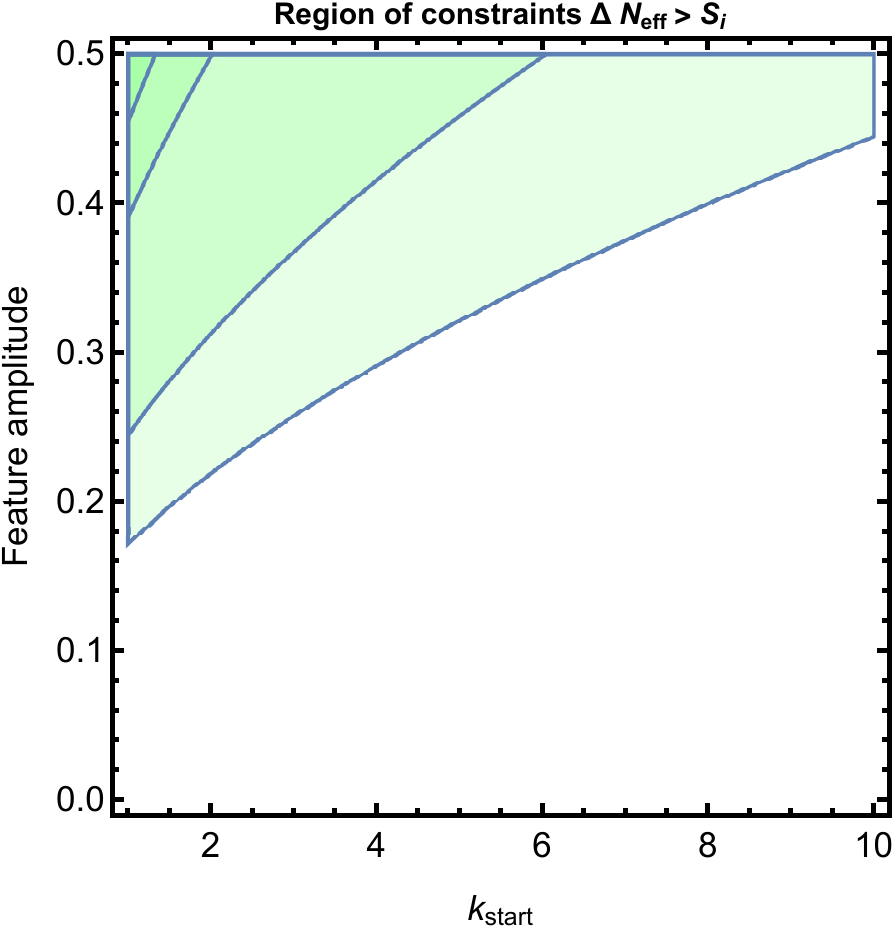}\\\includegraphics[width=0.5\textwidth]{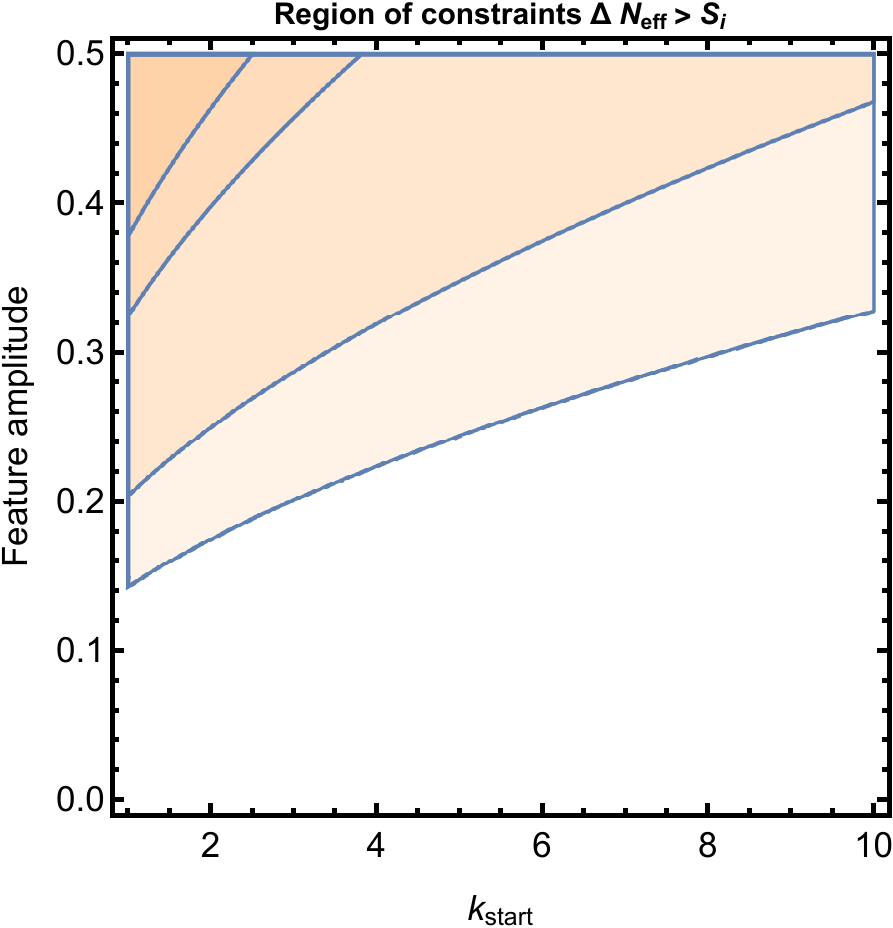}\includegraphics[width=0.5\textwidth]{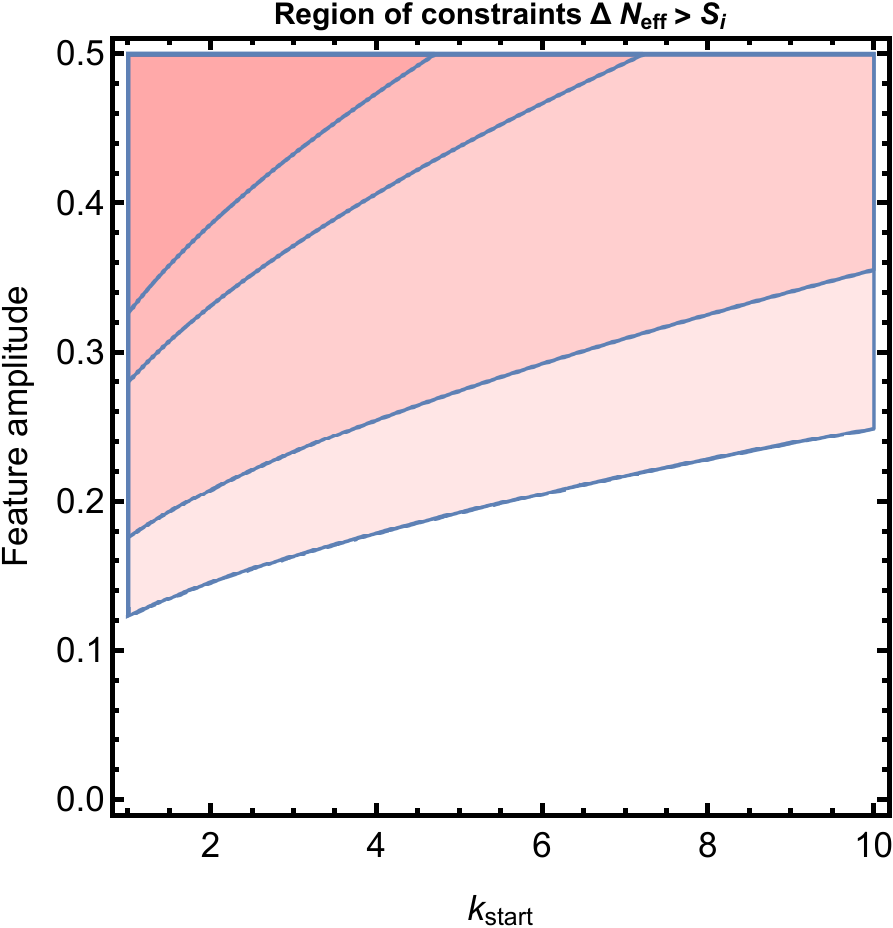}
\caption{A square feature over the standard $P_S$ where the UV cutoff is at $k_{UV}/k_0=10^{30}$ and the horizontal axis is the initial $k$ where the feature starts. For reference $k_{start}=1$ means $k_{start}/k_0=20$ . The different panels correspond to the thickness of the feature in k-space, from $\Delta k=1$ (upper left panel) through $\Delta k \in \{2,4\}$ to $\Delta k =8$ (bottom right panel).  As can be seen from the graph, the 'thickness' of the square feature is dominant over the $\Delta N_{eff}$ as recovered from the different experimental stages.}
\label{fig:feature_ira}
\end{center}
\end{figure}

Let us now consider the possibility if a brief period of particle production during inflation. Such a scenario produces a spectrum with a bump at the relevant wave number, \eqref{eq:bump}. In such a case, $N_{eff}$ is sensitive to the amplitude of the particle production, but insensitive to the wavenumber, i.e. the e-fold of inflation where it occurred. The only exception being if the bump is at the $k_i\simeq k_{UV}$ scale. Since the standard spectrum gives a negligible contribution to $N_{eff}$, the constraint is basically the integral over the feature. One can actually perform the integral analytically, and the leading term is given by the following:

\be
I\equiv \int^{k_{UV}} d \ln k P_S(k)^2\simeq \left(\frac{e}{3}\right)^3B^2\simeq0.74B^2
\ee
which corresponds to 
\be
B<0.55, \, current; \quad B<0.16, \, S4
\ee
Hence, it is not placing strong bounds on single particle production events. An example of a square feature was numerically integrated and the results are presented in Figure \ref{fig:feature_ira}

\subsection{Scalar spectrum with a step feature}
\begin{figure}[!h]
\begin{center}
\includegraphics[width=0.75\textwidth]{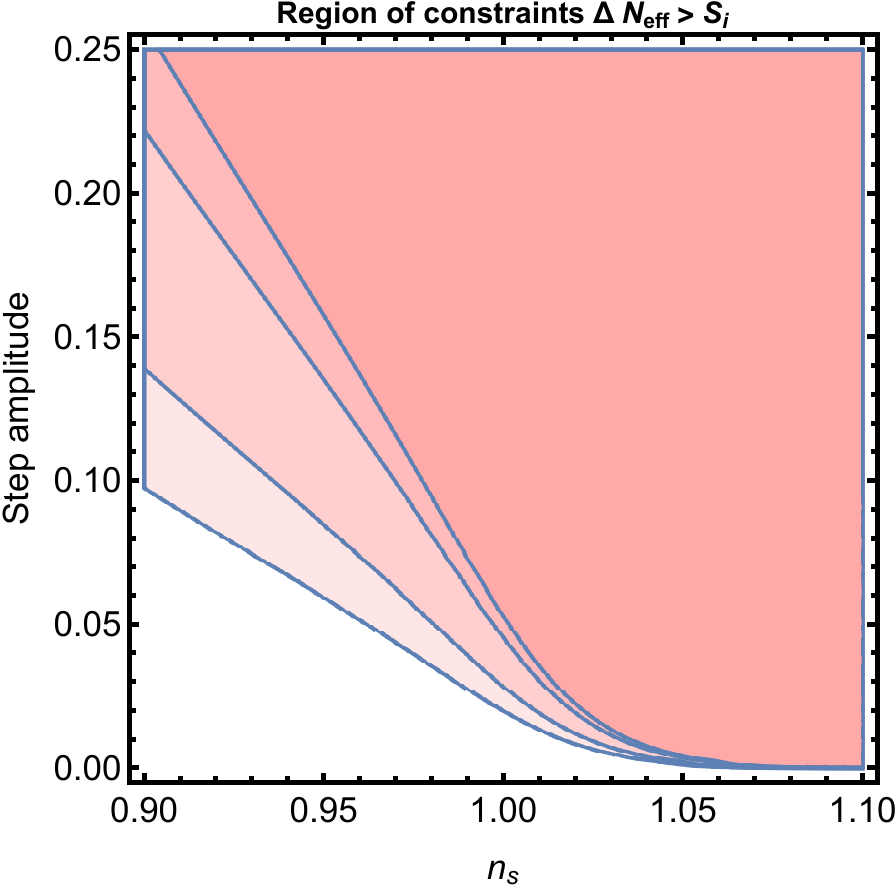}
\caption{The step function is located at $k_i=20k_0$, and what is shown are the constraints for different experimental stages (S1-S4), at the step amplitude vs. underlying $n_s$. }
\label{fig:step_ira}
\end{center}
\end{figure}

A spectrum with a step can occur for instance if there are several periods of inflation. To limit the number of free parameters, we assumed that only the amplitude has changed while the spectral index remains the same before and after the step \eqref{eq:step}. 
\be
I\equiv \int^{k_{UV}} d\ln k \, P_S^2\simeq \frac{B^2}{2(n_s-1)} \left(\left(\frac{k_{UV}}{k_0}\right)^{2 (n_s-1)}-\left(\frac{k_i}{k_0}\right)^{2(n_s-1)}\right)
\ee
Since the spectral index is still slightly red, the constraint here is weak. Basically $B\lesssim 10^{-2}-10^{-1}$ depending on wavenumber where the step occurs ($k_i$). A prototypical example is depicted in Figure \ref{fig:step_ira}.

\subsection{Scalar spectrum with a bend feature}

\begin{figure}[!h]
\begin{center}
\includegraphics[width=0.5\textwidth]{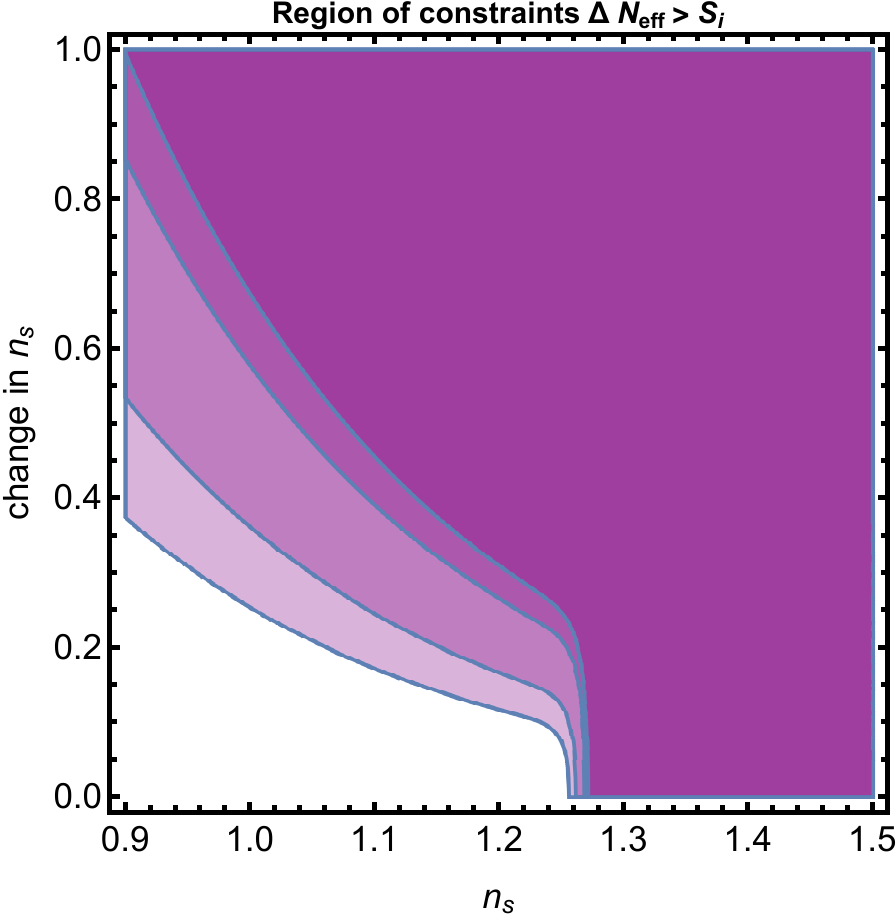}\includegraphics[width=0.5\textwidth]{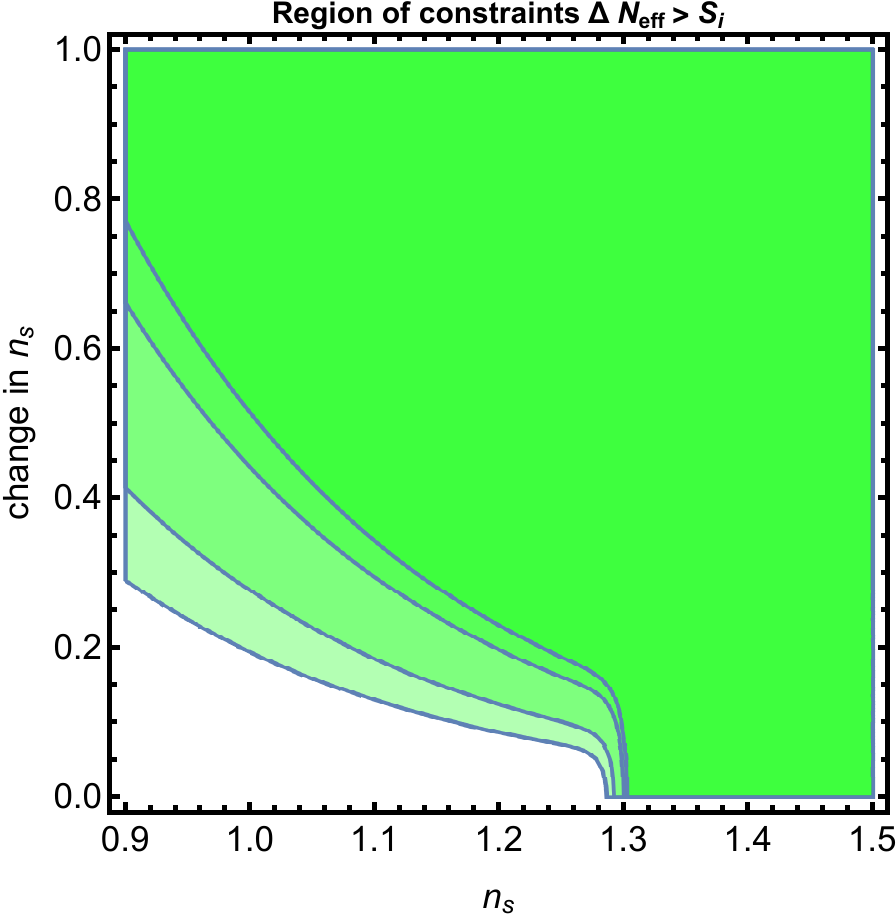}\\
\includegraphics[width=0.5\textwidth]{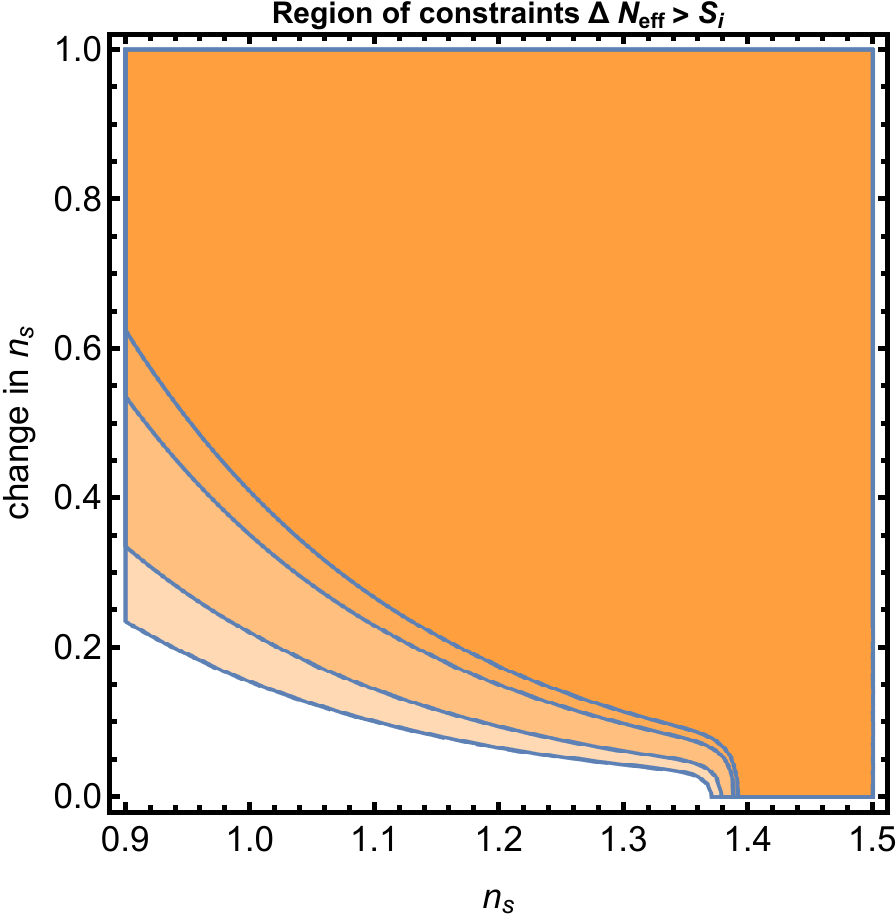}\includegraphics[width=0.5\textwidth]{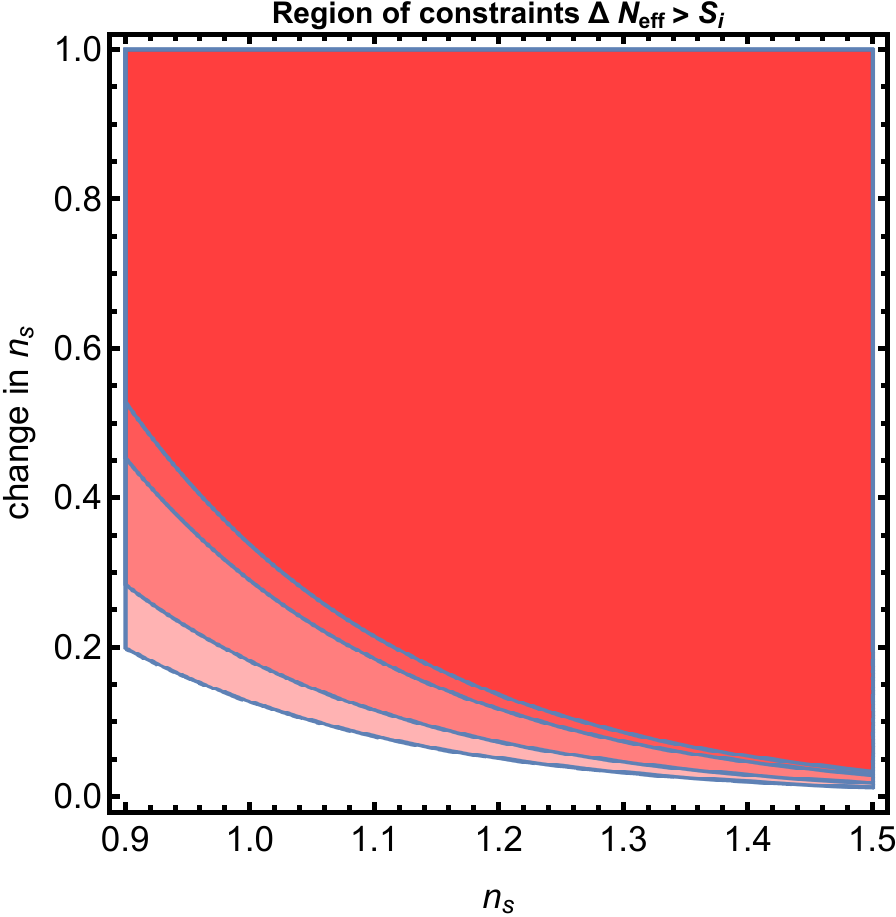}
\caption{ The bend is located at $k_i/k_0=20$, and what is shown is the constraints for different experimental stages (S1-S4), at the bend amplitude vs. underlying $n_s$. This is done for $k_{UV}/k_0 \in \{10^{21},10^{24}\}$ (upper row, left to right) and $k_{UV}/k_0 \in \{10^{27},10^{30}\}$ bottom, left to right.}
\label{fig:bend_ira}
\end{center}
\end{figure}

The bend parameterization provides a reasonable approximation to models with non-monotonic slow-roll parameter $\epsilon$, such that $r\gtrsim 0.01$ while the field excursion is still small, $\Delta \phi\leq 1$ \cite{BenDayan:2009kv,Wolfson:2016vyx,Wolfson:2018lel}.
\be
I\equiv \int^{k_{UV}} d\ln k \, P_S^2\simeq \frac{A_s^2}{2(n_s+n_s^*-2)}\frac{k_{UV}^{2(n_s+n_s^*-2)}}{k_0^{2(n_s-1)}k_i^{2(n_s^*-1)}}\; .
\ee
Since the integral is dominated by $k_{UV}$ we get a similar behavior to a constant $n_s>1$ depending on $k_i$ and $k_{UV}$. 
An example of the results for 
$\frac{k_i}{k_0}=20$ is presented in Figure \ref{fig:bend_ira}.
 
 \section{Likelihood Analysis \label{sec:Likelihood}}
  \begin{figure}
\includegraphics[width=7.3cm,height=73mm]{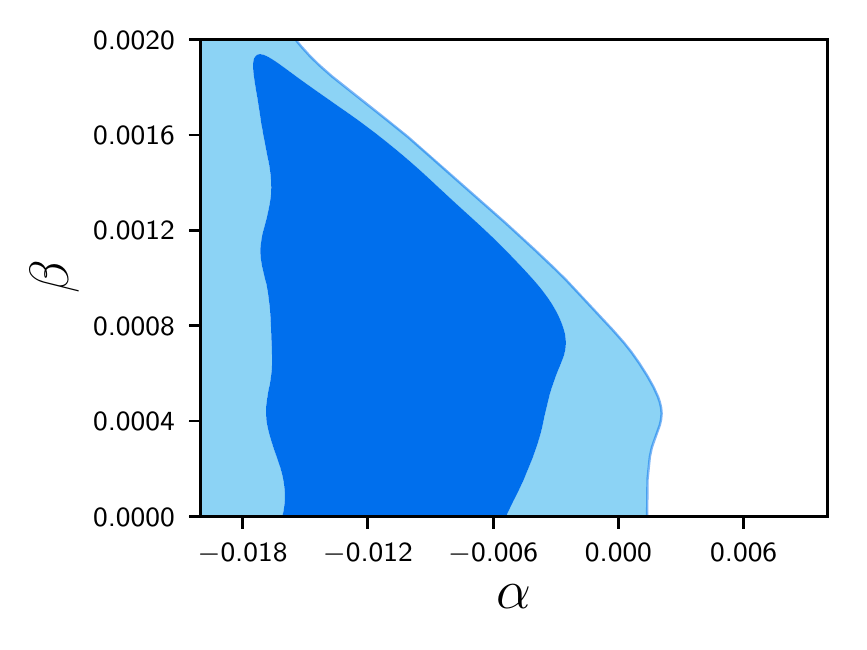} 
\includegraphics[width=7.3cm,height=73mm]{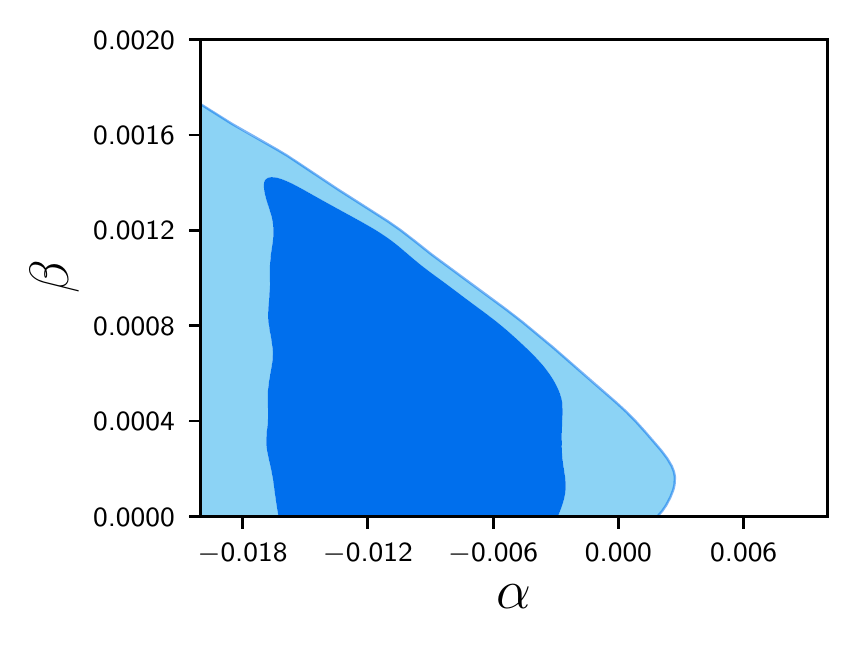}\\ 
\includegraphics[width=7.3cm,height=73mm]{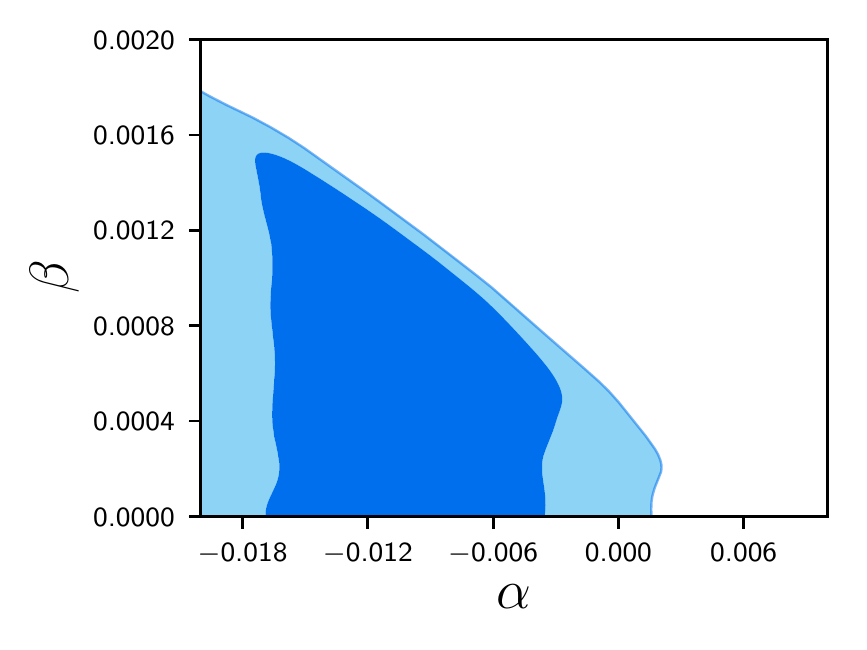}
\includegraphics[width=7.3cm,height=73mm]{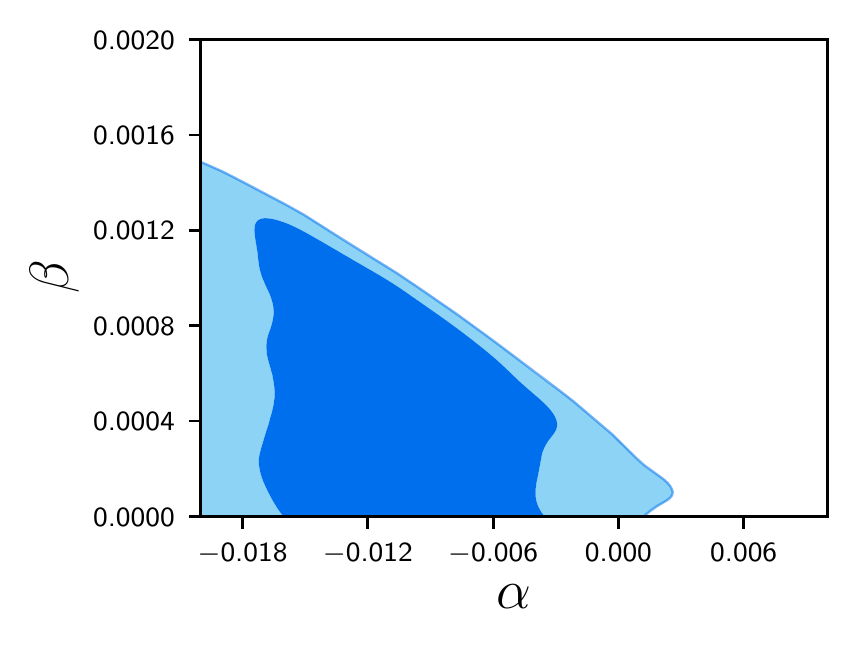}
\caption{Top panel: 68\% and 95\% confidence level contours of
$\alpha,\beta$
taking into account the bound on $2\Delta N_{eff}\leq 0.46 $ for $n_s(k_0)=0.95$ and  $k_{UV}/k_0=10^{21}$ (top left), $k_{UV}/k_0=10^{24}$ (top right).  Bottom panel: Same as top panel with $n_s(k_0)=0.97$.  S4 is expected to improve the constraints on $N_{eff}$ by an order of magnitude to $2\Delta N_{eff}\leq 0.04$.} 
\label{fig:likelihood}
\end{figure}
  The analysis in the previous section used \eqref{eq:bound} in a strict mathematical sense. However, a proper estimation of cosmological parameters, requires a likelihood analysis allowing several parameters to vary with proper priors.
 We have seen that the bound is most useful in constraining the running $\alpha$ and the running of running $\beta$. We therefore ran a CosmoMC analysis allowing the variation of $\alpha, \beta$ and
 we considered spectral tilts $n_s=0.95,0.97$ and various cut-offs $k_{UV}/k_0=10^{21}, 10^{24}$ corresponding roughly to two decades beyond the LIGO band and $60$ e-folds of inflation respectively.
 
 Our likelihood analysis uses data from BICEP2, Planck, HST, BAO and the KECK/Planck cross correlation analysis \cite{Akrami:2018odb,Ade:2018gkx,Array:2015xqh}. In addition to the base CosmoMC software distribution \cite{Lewis:2002ah}, we created an additional likelihood module to calculate $\Delta N_{eff}$ as a function of our running parameters,
 \beq
 \Delta N_{eff}(\alpha, \beta) = \left(3.046 + \frac{8}{7}\left(\frac{11}{4}\right)^{4/3}\right)\frac{A_{GW}^{(2)}}{36}\int^{k_{UV}}d\ln k\ P_{S,run}^2(k;\alpha,\beta),
 \eeq
 where $P_{S,run}$ is the one defined in \eqref{eq:run}.
 In the absence of a closed form solution to the integral for $\Delta N_{eff}$, we precomputed a grid of values for different inputs of $\alpha, \beta$ that are then used in the MCMC analysis. The likelihood plots corresponding to each pair of $(n_s(k_0),k_{UV}/k_0)$ are plotted in Figure \ref{fig:likelihood}. They are the results of running our modified version of CosmoMC with its own precomputed grid of $N_{eff}$ values. We use spacings of $\Delta\alpha=4\times10^{-4},\Delta\beta=2\times10^{-5}$ and compute $N_{eff}(\alpha,\beta)$ using bilinear interpolation for generic values of $\alpha,\beta$.
 
 It is clear that both parameters are severely constrained with $\alpha, \beta<0.002$. This is over an order of magnitude improvement compared to Planck bounds, and in accord with standard slow-roll predictions of $\alpha\sim(n_s-1)^2$ and $\beta\sim (n_s-1)^3$. 
 Taken at face value, the results again disfavor PBH dark matter models that require $\beta>0.002$  \cite{Munoz:2016owz}. 
 On the more general level, while the bound does not preclude features in the primordial power spectrum, it certainly weakens the case for a continuous feature while localized features in some small domain of wave-numbers $k$ are still plausible.
 Combining the analysis of this section and the previous one, we can place upper limits on linear combinations of $\alpha$ and $\beta$ for each distribution. Table~\ref{tab:upperlimits} summarizes the 68\% 95\% and 99.7\% upper limits for each distribution shown in Figure~\ref{fig:likelihood}. The most conservative analysis ($3\sigma$) gives the following bound $\beta + 0.074\ \alpha<8.6\times 10^{-4}$.
  It would be interesting to include the LIGO data in a future likelihood analysis, potentially strengthening these bounds.

\begin{table}
\begin{center}
\caption{Upper Limits on combinations of $\alpha$ and $\beta$ obtained from our CosmoMC likelihood distributions}
\label{tab:upperlimits}
\begin{tabular}{c|c|c|ccc}
$n_s(k_0)$  & $k_{UV}/k_0$ & parameter & 68\% u.l. & 95\% u.l. & 99.7\% u.l. \\
\hline
0.95 & $10^{21}$ & $\beta + 0.074\ \alpha$ & $3.8\times 10^{-4}$ & $7.7\times 10^{-4}$ & $8.6\times 10^{-4}$ \\
0.95 & $10^{24}$ & $\beta + 0.063\ \alpha$ & $2.1\times 10^{-4}$ & $5.0\times 10^{-4}$ & $5.5\times 10^{-4}$ \\
0.97 & $10^{21}$ & $\beta + 0.067\ \alpha$ & $2.1\times 10^{-4}$ & $5.1\times 10^{-4}$ & $5.7\times 10^{-4}$ \\
0.97 & $10^{24}$ & $\beta + 0.059\ \alpha$ & $1.2\times 10^{-4}$ & $3.6\times 10^{-4}$ & $4.1\times 10^{-4}$ \\
\end{tabular}
\end{center}
\end{table}

 \section{Constraints and forecast from other gravitational waves experiments\label{sec:constraints}}
 The absence of stochastic GW at LIGO and PTA scales allows us to place direct constraints on the fractional energy density stored in GW produced from scalars, and hence on the primordial power spectrum. For this we do not need to integrate over the spectrum, as we can place direct bounds on the GW or scalar amplitude at each wavenumber. Notice that these bounds do not depend on parameterization, but rather a strong bound on the amplitude of the power spectrum at these scales.
Using the expressions \eqref{eq:o2gw},\eqref{eq:fo2gw} and the fact that $a_{eq}/a_{today}\simeq 1/3400$  and $A_{GW}^{(2)} \simeq 10$ we have
\be
\Omega^{(2)}_{GW}(k, \eta_0)=\frac{P_S^2(k)}{340}, \quad k>58\, k_{eq}
\ee 
Given the bounds from LIGO and PTA, as well as future constraints from LISA, SKA and aLIGO we can constrain the primordial power spectrum, $P_S$.
Using \eqref{eq:ftok} the current LIGO and PTA measurement are specified in Table \ref{tablecurrent}. Notice that these bounds are already better than ones obtained by the absence of primordial black holes.  Forecasted constraints assuming no detection appear in Table \ref{tableforecast}. 
\begin{table}
\begin{center}
\caption{Current constraints on the primordial scalar power spectrum from current GW observations}
\label{tablecurrent}
\begin{tabular}{c|c|c}
$\mathrm{Experiment}$  & $A_s$ & $ k\, \mathrm{ (Mpc^{-1})}$ \\
\hline
LIGO & $<7.6 \times 10^{-3}$ & $3\times 10^{16}-1.3 \times 10^{17}\mathrm{Mpc^{-1}}$ \cr
PTA & $<5.8 \times 10^{-4}$ & $\sim 1.5 \times 10^8\mathrm{ Mpc^{-1}}$
\end{tabular}
\end{center}
\end{table}

\begin{figure}
      \centering
      \includegraphics[width=0.9\textwidth]{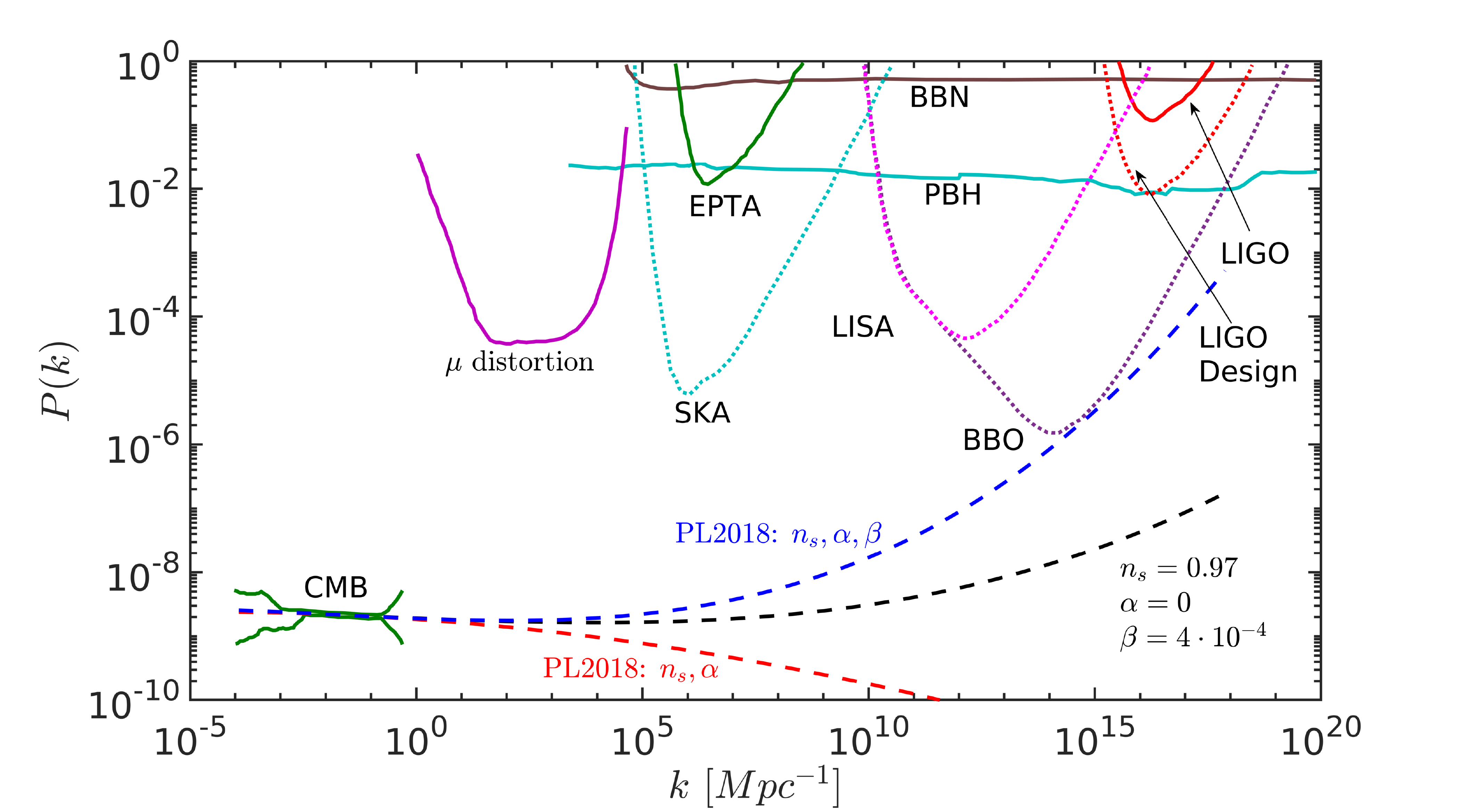}
      \caption{The map of the scalar power spectrum constraints by current and expected data. The dashed lines represent PPS allowed models. The red dash is the Planck 2018 analysis with $n_s=0.9641,\alpha=-0.0045$, where the blue dash is $n_s=0.9647,\alpha=0.0011,\beta=0.009$. The black dash is one of the least constrained models that are still allowed by our analysis with $n_s=0.97,\alpha=0,\beta=4\cdot 10^{-4}$ }
      \label{fig:PS_Constraints}
  \end{figure}

\begin{table}[h!]
\begin{center}
\caption{Forecasted constraints on the primordial scalar power spectrum from future GW observations}
\label{tableforecast}
\begin{tabular}{c|c|c}
Experiment  & $A_s$ & $ k\, \mathrm{(Mpc^{-1})}$ \\
\hline
aLIGO & $<5.8 \times 10^{-4}$ & $3\times 10^{16}-1.3 \times 10^{17}$  \\
SKA-PTA & $< 1.8 \times 10^{-5}$ & $\sim 1.5 \times10^8$  \\
LISA & $<5.8 \times 10^{-6}$ & $1.5 \times10^{12}- 1.5 \times 10^{13}$ \\
\end{tabular}
\end{center}
\end{table}
These results combined with a compilation of other probes of the power spectrum are given in Figure \ref{fig:PS_Constraints}, where we used \cite{Byrnes:2018txb,Inomata:2018epa}. In the figure there are also a few examples of possible spectra. The dashed red one corresponding to negative running $\alpha$, and most likely spectrum parameterized in Planck 2018, hence no detection is expected in any future probe. The dashed blue corresponds to a detectable spectrum in LIGO in the future, also most likely from Planck 2018. However it is in discord with the $N_{eff}$ bound, assuming integration up to at least $k_{UV}/k_0=10^{21}$. Finally, the black dashed curve corresponds to positive $\beta$, but does not violate the $N_{eff}$ bound, and cannot be observed by planned experiments, even considering second order contributions. 
The details of these models are given in Table \ref{tab:Shown_models}.
\begin{table}[!h]
    \centering
      \begin{tabular}{||c|c|c|c||}
  \hline
       Analysis & $n_s$ & $\alpha$ & $\beta$\\
       \hline
       PL2018, $n_s,\alpha$&0.9641 & -0.0045& N/A \\
       PL2018 $n_s,\alpha,\beta$&0.9647 &0.0011 &0.009 \\
       This paper &0.97 &0 &0.0004 \\
       \hline
  \end{tabular}
    \caption{Most likely scalar spectra form the Planck 2018 analysis with running, and possible running of running. Additionally the third row contains one of the models recovered from this analysis, which is on the cusp of the $68\%$ CL once the $N_{eff}$ bound is included. This shows that our analysis further constrains the scalar spectrum while compliant with the Planck data. }
    \label{tab:Shown_models}
\end{table}

Regarding the specific parameterization of $\alpha $ and $\beta$ the LIGO constraint is again the strongest yielding $\beta=0,\alpha\leq 0.018$ and $\alpha=0,\beta\leq 0.0013$ from current constraints, and 
$\beta=0,\alpha\leq 0.015$ and $\alpha=0,\beta\leq 0.0011$ with design sensitivity. 
However the integral constraint from $N_{eff}$ is stronger for higher $k_{UV}$.
The PTA bound on $\alpha,\beta$ is also of relevance, as it involves extrapolation only up to $k\sim 10^8 \, Mpc^{-1}$ rather than $k\sim 10^{16}\, Mpc^{-1}$. In such case $\beta=0,\alpha<0.056$ and $\alpha=0,\beta<0.0077$. Future constraints will improve the bounds to $\beta=0,\alpha<0.041$ and $\alpha=0,\beta<0.0057$. This is an independent bound on  $\alpha,\beta$ and is an improvement compared to existing bounds on the running of running, $\beta$.
\section{Discussion and Conclusions}
 The joint analysis of CMB, PTA and LI experiments that involves scales separated by orders of magnitude shows a great promise in unravelling the mysteries of the Early Universe. We have taken a modest step towards a joint analysis by considering the constraining power of these upcoming experiments on the scalar primordial power spectrum.  Since a scalar primordial power spectrum has been observed, an inevitable consequence is the existence of a tensor power spectrum sourced from the interaction between scalar and tensor fluctuations at second order. This sourced tensor spectrum exists independently of the Early Universe paradigm that reigned, and is constrained in principle by CMB, PTA and LI observations. Considering the basic prediction of $r\sim 10^{-6}$ on CMB scales, our analysis shows that a cosmic variance limited CMB experiment with partial sky coverage and no delensing may be able to detect it. This calls for a more accurate estimate of the theoretical prediction, as well as a more detailed analysis of the systematics of such experiment. 
 
 Furthermore, this sourced spectrum is a function of the primordial scalar spectrum. Hence by considering CMB, PTA and LI observations and the relation between the scalar and sourced tensor spectrum, we can constrain the scalar spectrum in a way that has not been considered before and on length scales inaccessible to known probes.
 
We have demonstrated that considering the contribution of the sourced tensor spectrum to $N_{eff}$ yields an integral bound on the primordial scalar spectrum. The strength of the bound is parameterization dependent. Barring additional features in the spectrum, it forces the running $\alpha$ to conform to standard slow-roll results of $\alpha\sim (n_s-1)^2$, that is an order of magnitude better than primary CMB constraints. Moreover, the running of running is further constrained to be $\beta<0.002\lesssim (n_s-1)^2$.  S4 experiments are expected to improve these bounds.
Finally, we derived a direct bound on the amplitude of the primordial scalar spectrum using LIGO and PTA current measurements, as well as forecasts for future measurements.
This direct bound is, to our knowledge, the best bound on these scales, and improves bounds based on the absence of primordial black holes by more than an order of magnitude. 
The results presented here strengthen the case of ``vanilla type" inflation without additional features. 

\section*{Acknowledgements}
The research of IW was supported by the Israel Science Foundation grant no. 1294/16. IBD, BK and DL gratefully acknowledge support from the Ax Center for Experimental Cosmology at UC San Diego. This is not an official Simons Observatory  Collaboration paper.

\end{document}